# A Crater Chronology for the Jupiter's Trojan Asteroids

S. Marchi[1], D. Nesvorný[1], D. Vokrouhlický[2], W.F. Bottke[1], H. Levison[1]

1: Department of Space Studies, Southwest Research Institute, 1050 Walnut Street, Suite 300, Boulder, CO 80302
2: Institute of Astronomy, Charles University, V Holešovičkách 2, CZ–18000 Prague 8, Czech Republic

**Abstract.** We present a new crater chronology for Jupiter's Trojan asteroids. This tool can be used to interpret the collisional history of the bodies observed by NASA's Lucy mission. The Lucy mission will visit a total of six Trojan asteroids: Eurybates, Polymele, Orus, Leucus, and the near equal mass binary Patroclus-Menoetius. In addition, Eurybates and Polymele each have a small satellite. Here we present a prediction of Trojan cratering based on current models of how the Solar System and the objects themselves evolved. We give particular emphasis to the time lapsed since their implantation into stable regions near Jupiter's Lagrangian $L_4$ and $L_5$ points.

We find that cratering on Trojans is generally dominated by mutual collisions, with the exception of a short period of time (~10 Myr) after implantation, in which cometary impacts may have been significant. For adopted crater scaling laws, we find that the overall spatial density of craters on Trojans is significantly lower than that of Main Belt asteroids on surfaces with similar formation ages. We also discuss specific predictions for similar-sized Eurybates and Orus, and the binary system Patroclus-Menoetius.

**1. Introduction.**

Crater chronologies are an important tool for assessing the evolution of solid surfaces across the Solar System. This method has been extensively applied to study the geology and collisional evolution of near-Earth and Main Belt asteroids (MBAs) visited by spacecraft (e.g., for a review see Marchi et al. 2015). These studies benefitted from a relatively well understood collisional evolution of these populations (e.g., Bottke et al. 2015; 2020). The selection of NASA's Lucy mission to flyby eight Trojan asteroids of Jupiter (hereafter Trojans) raised the need to develop a crater chronology specific for these objects.

To this goal, we rely on numerical simulations to define the present Trojan impact environment, as well as how the impact rate has changed since their formation. It is generally accepted that the Trojan asteroids were captured in Jupiter's $L_4$ and $L_5$ Lagrangian points early in Solar System history during the migration of the giant planets (Morbidelli et al. 2005; Nesvorny et al. 2013). The precise nature of the Trojan capture mechanism has important consequences for how the bodies experienced collisional evolution. An outstanding issue is the timing of their capture, generally believed to have occurred within 100 Myr since Solar System formation (e.g., Nesvorny 2018). In the leading models for Trojan formation, the objects formed in the outer planetesimal disk between ~20-30 au. The initial planetesimal population was thought to have been massive, at least ~15-20 Earth masses. It was dispersed by an outwardly migrating Neptune that drove planetesimals both into the giant planet zone and into the Kuiper belt and scattered disk. A small fraction ($<10^{-6}$) of the inward-scattered planetesimals were likely captured as Jupiter Trojans via gravitational interactions with the giant planets by chaotic capture mechanisms (Morbidelli et al. 2005) or jump capture mechanisms (Nesvorny et al. 2013). These capture models explain the number of objects in the Trojan population and their orbital distribution, and may also explain the spectroscopic similarity between Jupiter Trojans and Kuiper Belt objects (Emery et al. 2015). Alternative formation models are discussed in Emery et al. (2015).

Here we limit our attention to the Trojans captured by the jump capture mechanism discussed by Nesvorny et al. (2013). There are three stages in this model: (i) planetesimals form and reside in the outer disk beyond ~15-20 au, where they undergo collisions with other nearby planetesimals, (ii) planetesimals that are scattered inward by Neptune and reach the vicinity of Jupiter are captured as Trojans, and iii) Jupiter Trojans captured on dynamically unstable orbits decay over time, while all Trojans collide with themselves and with objects from other populations (e.g., comets). The duration of stage (i) is unknown, but various constraints suggest it lasted ~10-100 Myr (Nesvorny et al. 2018; Riberio et al. 2020).

The degree of collisional evolution taking place during stage (i) must have been intense given the large estimated mass of the outer disk. In comparison, stage (ii) was probably brief (< 5 Myr), though its importance for impact cratering may be non-trivial, given that the proto-Trojans were surrounded by a large unstable population as they approached Jupiter. As the populations destabilized by Neptune's migration became dynamically depleted over time, the cratering rate during stage (iii) on the captured Trojans reached a relatively low-level state compared to (i). However, the Trojan collisions with one another has been going on for 4.5 Gyr, allowing this phase to produce substantial numbers of craters.

While significant uncertainty remains about the early evolution of Trojans, we derive here a model for their collisional evolution with the scope to provide a general reference frame for the cratering histories of the Lucy targets. We focus on stage (iii), partly because craters formed during that time are superposed on older craters, but also because that is where we have more confidence in the impact flux modeling. Any excess of crater populations over those expected from stage (iii) should be attributed to stages (i) and (ii), as we will discuss later.

**2. Trojan's impact rates.**

At present, Trojan asteroids are primarily impacted by four distinct populations: Trojans themselves, Hilda asteroids, Thule asteroids, and comets with perihelion distances below 5.2 au. Marzari et al. (1996) and Dell'Oro et al. (1998) computed the impact velocity ($v$) and rate of impacts per unit time per unit surface called intrinsic probability of collisions ($P_i$) for the present populations of Trojans-Trojans, Trojans-Hildas and Trojans-Thules (see Table 1). For these impact populations, we adopt the present-day $P_i$ and mean impact velocity ($v_m$) from prior models (Marzari et al. 1996), which were computed using orbits of a set of numbered Trojan, Hilda and Thule asteroids.

We note that Trojans-Trojans collisions have $P_i$ that is more than a factor of 2 higher than collisions among MBAs (the latter is $2.9 \times 10^{-18}$ yr$^{-1}$ km$^{-2}$; based on MBAs with $d > 50$ km; Bottke et al. 2020), while the impact velocity distribution is similar. This is counter intuitive as both $P_i$ and $v_m$ are expected to decrease with heliocentric distance. This effect is compensated by the relatively small volume occupied by each Trojan cloud compared to MBAs.

To estimate Comets-Trojans collision probabilities, we rely on the dynamical model of Nesvorny et al. (2017). We monitored cometary impacts on a target body that was placed at 5.2 au, the average Trojan semimajor axis value, with zero eccentricity and inclination. The zero eccentricity and inclination for the target body results in a negligible error on the computed collisional probabilities (<10 %), given the wide orbital distribution of comets. We used the numerical integration results from Nesvorny et al. (2017) to track impacts with the target body in the earliest evolutionary phase ($t \gtrsim 4.3$ Ga; first 200 Myr of stage (iii)), in which there are good statistics of cometary projectiles in the simulations. For later times ($t \lesssim 1$ Ga), we used cloning techniques to increase comet population statistics. All comets that reach a heliocentric distance less than 9 au are cloned. The cloning is done by a small perturbation of the velocity vector (~$10^{-6}$ relative to vector's magnitude). We used 100 clones in these runs. Our calculated intrinsic collision probability between comets and Jupiter Trojans at the present time is $P_i = 0.96 \times 10^{-18}$ yr$^{-1}$ km$^{-2}$. We also calculated $P_i$ in 100 Myr intervals in the last Gyr and also in the first 200 Myr after the start of the simulation. We find that $P_i$ only varies by a factor of ~2 at most.

The impact velocity distributions from these calculations for Trojans-Trojans and Comets-Trojans are shown in Figure 1, with the mean values provided in Table 1.

|                        | $P_i$ (yr$^{-1}$ km$^{-2}$) | $v_m$ (km s$^{-1}$) |
|---|---|---|
| Trojans-Trojans        | $(7.00 \pm 0.10) \times 10^{-18}$ | 4.6  (1) |
| (Hildas+Thules)-Trojans | $(0.24 \pm 0.10) \times 10^{-18}$ | 4.2  (1) |
| Comets-Trojans         | $(0.96 \pm 0.10) \times 10^{-18}$ | 6.0  (2) |

Table 1. Intrinsic collisional probabilities and average impact velocities used in this work. (1) Marzari et al. (1996). Note formal uncertainties for $P_i$ are rounded to excess for Trojans-Trojans, while for Hildas+Thules we have estimated the uncertainty by comparing nominal $P_i$ variations for impacts with L$_4$ and L$_5$ bodies. (2) This work.

The present number of impacts per unit surface per unit time ($N_0$) is obtained using the equation:

$$N_0(>d) = P_i \times S(>d) / (4\pi)$$

where $S(>d)$ is the impactor cumulative size-frequency distribution (SFDs) and $d$ is projectile diameter. Our model Trojan and comet SFDs come from Bottke et al. (2023), who computed the combined collisional and dynamical evolution of the primordial Kuiper Belt, destabilized objects ejected from the primordial Kuiper Belt that achieved giant planet-crossing orbits, and objects from the latter population that were captured as Jupiter Trojans. Here we restrict our impact calculations to a single Trojan cloud ($L_4$ and $L_5$ are assumed to have identical $S(>d)$; Figure 2). Further, we assume that the Jupiter family comets (JFCs) are the primary comet population that produce Trojan cratering (Zahnle et al. 2003). Our numerical simulations calculate the number of impacts with comets at a reference size of 10 km in diameter. We rescale this number to the present population of JFCs, estimating the present number of JFCs that are 10 km or larger to be ~25. This value is based on an estimate of the debiased JFC SFD by WISE (Bauer et al. 2017) that was fitted by collisional modeling results from Bottke et al. (2023). Our model shows that Oort cloud comets have a negligible contribution to Trojans craters (< 1% of the JFCs impacts).

The preset Hilda SFD is similar in slope to that of Trojans, but has fewer objects by a factor ~4-5 for $d > 10$ km (Terai and Yoshida 2018; Davis et al. 2002), while Thules are a factor of ~100 fewer in number than Hildas for $d > 10$ km (Broz and Vokrouhlicky 2008). Also, $P_i$ for (Hildas+Thules)-Trojans collisions is a factor of ~30 smaller than Trojans-Trojans (Table 1). Put together, we estimate that the present (Hildas+Thules)-Trojans impact flux is at least a factor of 100 smaller than for Trojans-Trojans collisions. For this reason, we assume that Hildas and Thules cratering is at present negligible. For completeness, we note that there exists an additional population of asteroids that can collide with Trojans, the so-called Hecuba-gap group (Broz and Vokrouhlicky 2008). This group is located further from Jupiter and therefore has lower $P_i$ with Trojans compared to Hildas. Moreover, Hecuba-gap asteroids are about a factor of ~15 fewer in number than Hildas for $d > 10$ km, and for these reasons they can be neglected.

We stress that formal uncertainties associated with the values listed in Table 1 are typically of a few %, thus negligible for the purposes of studying crater populations. However, a more realistic way to assess uncertainties is to compute $P_i$ using different orbital distributions for asteroids and comets, which goes beyond the scope of this work. Here we assume that $P_i$ remains constant over time (within a factor of 2, as noted above, for the most critical Comets-Trojans case), and study how the number of bodies in each impactor population changes over time.

In order to build a crater chronology, we need to know how the impact rate evolved over time. This can be assessed using models for the formation of Trojan asteroids. Here we use the above comet model to track their impacts onto Trojans over time, while using the Nesvorny et al (2013) model to compute the rate of Trojans-Trojans impacts over time.

The calculated Trojans-Trojans and Comets-Trojans cumulative number of impacts, $N_i(<t)$, is shown in Figure 3, and can be written as:

$$N_i(<t) = (4.5/b) \times [1 - (1- t/4.5)^b]$$

where $t$ is in Ga ($t = 0$ is the present), and $b = 0.82$ and $-1.1$ for Trojans-Trojans and Comets-Trojans impacts, respectively. The total number of impacts is:

(1)  $N(<t, >d) = N_0(>d) \times N_i(<t) = [P_i \times S(>d) / (4\pi)] \times (4.5/b) \times [1 - (1- t/4.5)^b]$

As a further test for JFC impact rates, we calculated the current number of impacts on Jupiter. Our work yielded $3 \times 10^{-5}$ $d > 10$ km impacts on Jupiter per year. This value compares well with $3.4 \times 10^{-5}$ $d > 10$ km impacts on Jupiter per year, and it will be discussed in more detail in a follow up study dedicated to this topic.

There are several caveats to these estimates. First, there are some differences between Trojan formation simulations with different initial conditions, and these differences become more important (a factor of ~2 fluctuation) when $t$ approaches the beginning of stage (iii); the chronology functions discussed above do not apply to stages (i) and (ii). Second, we fit the computed impacts of comets on Trojans over time with a power law. Third, given that we do not have good statistics for comet impacts in the $t = 1-4$ Ga interval, we are unable to verify the precise nature of their decay profile with time. So we adopt a power law and tie it to the simulation results for $t \gtrsim 4.3$ Ga and $t \lesssim 1$ Ga where numerical statistics are better.

Another important aspect of this problem concerns the past evolution of the Hilda population. We have shown that presently Hildas do not significantly contribute to Trojans' cratering. Dynamical models show that Hildas and Trojans have about the same implantation probability (Vokrouhlický et al 2016), and so it is expected their populations might have had a similar number of asteroids after implantation. Even under this circumstance, Hildas provide a smaller contribution to Trojans cratering by a factor of 30, as indicated by the present $P_i$ ratio, and such ratio is not expected to significantly change with time. A full model of the evolution of the Hilda population over time is left for future work.

In conclusion, we argue that our impact rates provide a reasonable baseline model for the cratering history of Trojans, but uncertainties remain.

**3. Cratering scaling laws.**

Here we use the so-called Pi-group scaling law (e.g., Hosapple and Housen 2007) that provides the transient crater diameter ($D_t$) as a function of impact conditions and material properties:

(2)  $$D_t = kd \left[ \frac{gd}{2v_\perp^2} \left(\frac{\rho}{\delta}\right)^{2\nu/\mu} + \left(\frac{Y}{\rho v_\perp^2}\right)^{(2+\mu)/2} \left(\frac{\rho}{\delta}\right)^{\nu(2+\mu)/\mu} \right]^{-\mu/(2+\mu)}$$

where $g$ is the target gravitational acceleration, $v_\perp$ is the perpendicular component of the impactor velocity, $d$ is the impactor diameter, $\delta$ is the projectile density, $\rho$ and $Y$ are the density and "cratering strength" of the target, $k$ and $\mu$ depend on the cohesion of the target material and $\nu$ on its porosity.

Equation (2) models the transient crater size resulting from the direct excavation and removal of target material. Craters on planetary surfaces are expected to undergo a post-formation modification phase in which fractured materials may flow back toward the center of the cavity. The final crater ($D_f$) is typically from 20% to 50% larger than the transient crater in rocky targets. We assume a similar relationship for the Trojans, and use 30% (Marchi et al. 2015). Figure 4 shows the computed crater size vs impactor size for average impact velocity for a generic 100-km Trojan asteroid. We implemented three different formulations of Eq. (2), namely a cohesive soil case for $Y = 10, 100$ kPa, and a porous case for $Y = 10$ kPa.

We stress, however, that the applicability of these relationships to Trojans is not well understood. The strength of the material is not known, and our assumptions are based on properties of known terrestrial assemblages. For instance, the strength of intact basalt (a hard rock) is of the order of 10 MPa, while dry alluvium (a cohesive soil) is of the order of 56 kPa. So our assumed $Y$ values cover a wide range of terrain properties from weak to moderate hard, but it is conceivable that Trojans might differ from our assumptions. Of note is the observation that the surface of small bodies (either comets or asteroids) are considered to be very low strength. For instance, Perry et al. (2022) found that the Bennu surface strength is less than 2 Pa, based on the ejecta pattern of a 70-m diameter crater. On the other hand, Ballouz et al. (2020) concluded that meter-sized boulders on Bennu have a strength 0.5-1.7 MPa. These strength values are inferred from meter-scale properties, and the applicability to craters from hundreds to several km in scale is not clear. All these caveats should be kept in mind while discussing predictions for Trojan's cratering in the next section.

**4. Results.**

From the inputs described in previous sections, we derive the Trojans cumulative distribution of craters as a function of time (the so-called model production functions, MPFs; Marchi et al. 2009, 2012, 2016). In Figure 5, we show example MPFs for 1 and 4 Ga, for both Trojans-Trojans and Comets-Trojans collisions. We find that Trojan self-cratering dominates crater production. However, older surfaces have proportionally larger fractions of comet craters.

Going back in time, the model predicts that at ~4.47 Ga, Trojans and comets contribute equally for similar scaling laws. We stress that these results should be taken with caution as impact rates are less well-understood at those early times, as discussed in the previous section. Nevertheless, these results raise an interesting point. Our modeling shows that the comet MPFs are shallower than Trojan MPFs in the size range 1-10 km craters. Craters of these sizes will be observed by Lucy, therefore crater SFDs on old surfaces could provide constraints on their origin (cometary vs asteroidal).

To further set expectations for what the Lucy mission might find, we use cratering from MBA Mathilde as a reference (Chapman et al. 1999). Among the MBAs visited by spacecraft at close range, Mathilde has bulk properties closest to the Trojans based on its C-type spectral class and a bulk density of ~1.3 g/cm$^3$. Mathilde's overall morphology is characterized by deep large craters that are consistent with it having a large porosity. From this comparison, we derive a few interesting conclusions (see Figs 6-7). If Trojans have fewer craters than Mathilde, these craters are very likely due to Trojan impactors. If Trojans have about the same number of craters observed on Mathilde, then there is a likely mixture of Trojan and comet craters going back in time to their capture. If, instead, Trojans have many more craters than Mathilde, it is possible that some craters formed prior to their implantation may yet be preserved on the surface.

We also note that prior work suggested that Mathilde's surface might be nearly saturated with craters (Marchi et al. 2015; Figs 6-7). If correct, this could limit how far back in time we are able to go looking at Trojan craters. Our expectation, however, is that Trojan cratering should be below saturation going back to the time of their implantation into $L_4$ and $L_5$, regardless of the crater scaling law adopted in our model (see Figs 6, 7).

Another interesting aspect of our work concerns the cratering history of the Lucy target Eurybates, the largest remnant of a collisional family. The age of the Eurybates family is not well constrained as it depends on the assumed strength for catastrophic disruption, $Q^*_D$ (Marschall et al. 2022). One way to constrain the family age is to calculate the probability that its satellite, Queta, survived since its formation; it is assumed that Queta formed at the same time as the Eurybates family itself. The Trojans SFD used here has a -2 cumulative slope for objects smaller than 10 km in diameter (Bottke et al. 2023). Using this SFD, and reasonable values of $Q^*_D$, the age of the family is likely to be considerably younger than 4.5 Ga (50% probability of being 1 Ga or less). If this is correct, then the Eurybates crater SFDs is solely due to Trojans-Trojans impacts, and should be significantly less cratered than other similarly sized Trojans, including Orus. For this reason, the comparison of crater SFDs between Eurybates and Orus may reveal whether the shape of the impactor population SFD has evolved over time.

We note that at present the Eurybates family accounts for about 10% of all $L_4$ Trojans larger than $d \sim 10$ km. The slope of the Eurybates family is steeper than the slope of the $L_4$ Trojans in the size range 15-20 km (Marschall et al. 2021). If such slope differences hold at smaller sizes, the Eurybates family could dominate the impactor population below $d \sim 2$ km. In this case, the crater SFDs could be steeper than our MPFs. On the other hand, the Eurybates family SFD in the size range 10-15 km has a slope similar to the $L_4$ Trojans, and the family would not be a significant source of smaller impactors compared to the background $L_4$ Trojans. These issues could be resolved by improving the Trojan SFD down to $d \sim 1-5$ km.

Finally, we note that cratering on Patroclus and Menoetius could inform us about the timing of their formation. Our expectation is that this binary system is primordial (e.g., Nesvorny et al. 2018). A level of cratering comparable or exceeding that of Mathilde would suggest the surfaces of the two worlds in the binary system are older than their implantation time in $L_5$, and may go all the way back to the origin of the binary itself.

### 5. Conclusions.

We derived a new crater chronology for the Jupiter's Trojan asteroids based on impact rates calculated from most updated dynamical models, and adopted impactor size-frequency distributions for various populations. The latter are based on updated collision evolution models for the Trojans asteroids and Kuiper Belt objects. We further adopted a range of crater scaling laws and material parameters. With these assumptions, we are able to calculate the collisional history of the Trojan asteroids since their implantation in $L_4$ and $L_5$. Our results for this model thereby provide a useful baseline reference for what the Lucy mission will observe, and give us insights into how craters can be used to develop a relative and absolute chronology for surface features on Trojan asteroids.

Using our model, we infer that Trojans may have fewer craters per unit surface area than MBAs with comparable absolute surface ages. We also find that Trojan asteroid surfaces may have not reached crater saturation, thus possibly preserving a record of the earliest cratering. Our results make predictions for Trojan cratering that can be used by the Lucy mission to estimate the approximate surface ages of the mission targets. Possible predictive uses of our model are as follows. First, craters on Eurybates may allow us to calculate the age of when its associated family formed. Second, the crater SFDs of similarly-sized Trojans like Eurybates and Orus may tell us how the impactor SFD has evolved with time. Third, Patroclus and Menoetius cratering may provide information of when the binary formed (e.g., primordial vs late formation).

**Acknowledgement.** The work of DV was partially funded by the grant 21-11058S of the Czech Science Foundation.

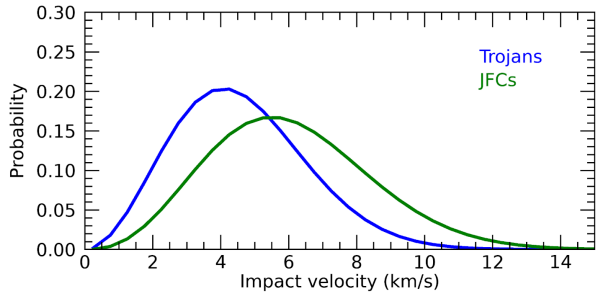

Figure 1. Computed impact velocity distributions for Trojans-Trojans and Comets-Trojans collisions, indicated respectively with Trojans and JFCs. For Trojans-Trojans, velocities are computed using a subset of numbered Trojan asteroids (Marzari et al. 1996). For Comets-Trojans, we considered a representative Trojan asteroid at 5.2 au (with zero inclination and eccentricity), and use an Opik-approximation to compute impact velocities with crossing comtes.

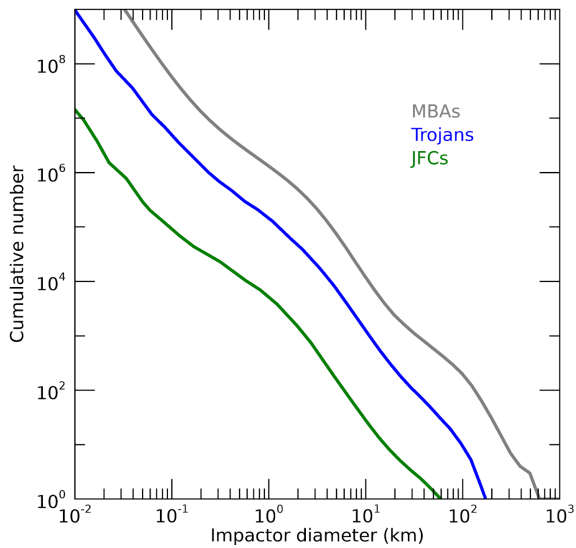

Figure 2. Population size-frequency distribution assumed in this work (see text for details). Note that Trojan SFD is only for one cloud, either $L_4$ or $L_5$. We also report MBAs for reference (taken from Bottke et al. 2020).

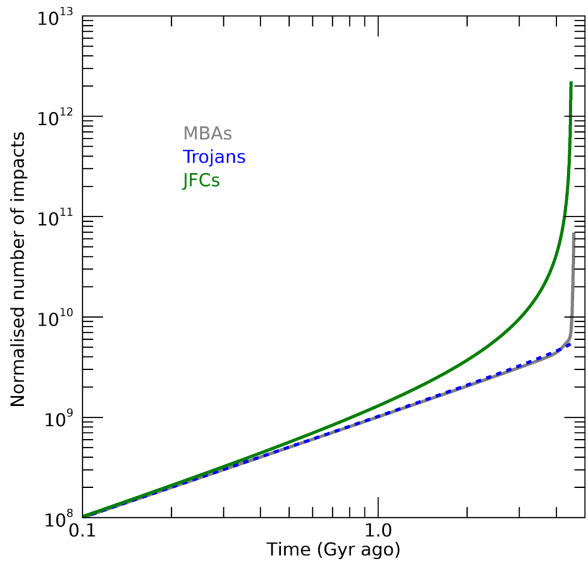

Figure 3. Chronology function used in this work (see text for details). For reference, a chronology for MBAs is also reported (O'Brien et al. 2014). All curves are normalized for a better comparison at the present time ($t = 10^{-9}$ Gyr ago). Because of the normalization, JFCs appear to have more impacts than the other chronologies, but in reality this is not the case as explained in the text.

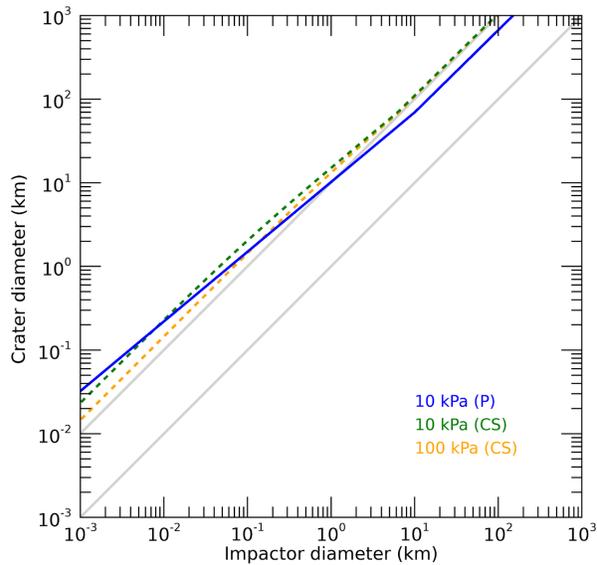

Figure 4. Calculated crater sizes on an assumed 100-km Trojan asteroid. Scaling laws are from Eq. (2) for cohesive soils (CS) and porous materials (P), cratering strength $Y$ as indicated (see text). Other parameters assumed are: density of target and projectile 1 g/cm$^3$. We assumed average impact velocity for Trojans-Trojans collisions (4.6 km/s; Table 1). Here we use a 45 deg impact angle. The material parameters are (Holsapple and Housen 2007): $\nu = 0.4$, $\mu = 0.41$, $k = 1.03$ (CS); and $\nu = 0.4$, $\mu = 0.4$, $k = 0.725$ (P). Gray lines indicate craters with diameters equal to 1x and 10x the impactor size.

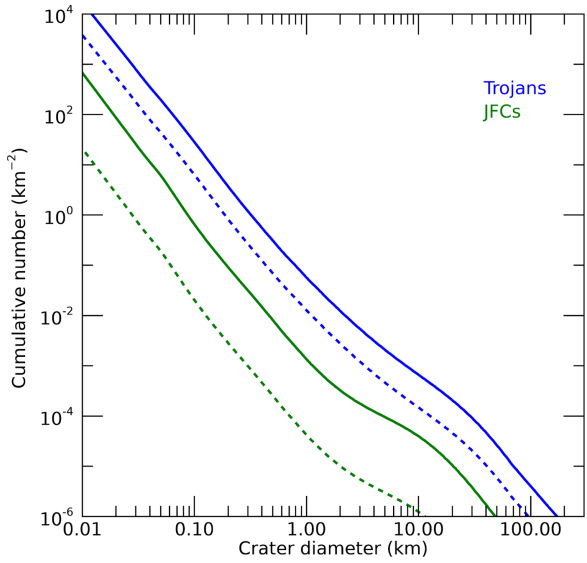

Figure 5. Trojan model production functions (MPFs) for cratering from Trojans and comets. Solid and dashed curves are for 4 and 1 Ga, respectively. This is for scaling law CS, $Y = 10$ kPa (see Fig. 4).

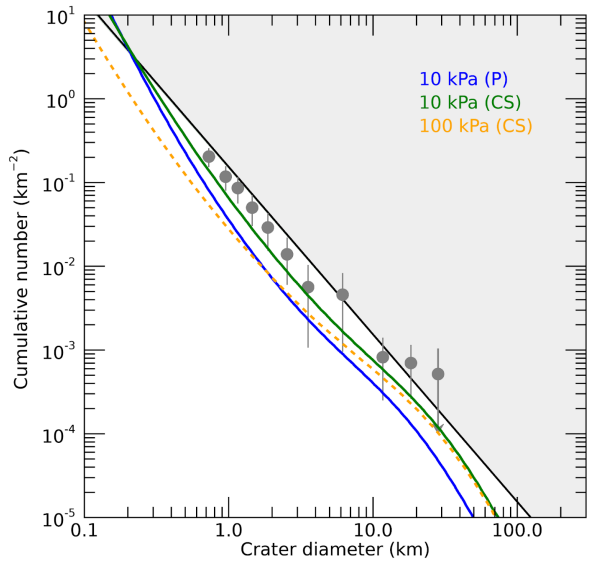

Figure 6. Trojans-Trojans MPFs for 4.4 Ga, compared with Mathilde cratering (see text). We report the results for three different scaling laws. The CS scaling with $Y = 10$ kPa strength comes closer to Mathilde cratering. Crater saturation is indicated by shaded area (black line corresponds to 10% of geometric saturation; see Marchi et al. 2015).

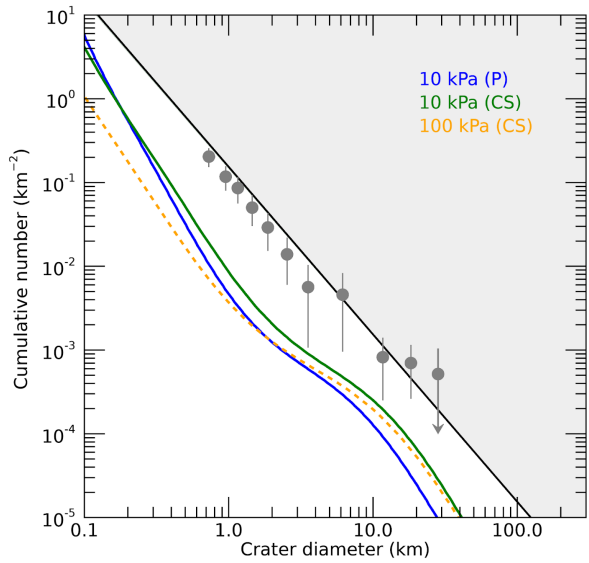

Figure 7. Comets-Trojans MPFs for 4.4 Ga, compared with Mathilde cratering (see text). Compare to Figure 6 for Trojans-Trojans cratering for the same crater scaling laws. Crater saturation is indicated by shaded area (black line corresponds to 10% of geometric saturation; see Marchi et al. 2015).